\documentclass[12pt]{article}
\usepackage{amsmath, amsthm}
\usepackage{amscd}
\usepackage{amssymb}
\usepackage{array}

\theoremstyle{definition}
\newtheorem{thm}{Theorem}[section]
\newtheorem{theorem}[thm]{Theorem}

\newtheorem{example}[thm]{Example}

\newtheorem{remark}[thm]{Remark}

\begin{document}

\newcommand{\id}{\relax{\rm 1\kern-.28em 1}}

\newcommand{\be}{\begin{equation}}
\newcommand{\ee}{\end{equation}}
\newcommand{\bea}{\begin{eqnarray}}
\newcommand{\eea}{\end{eqnarray}}
\newcommand{\bean}{\begin{eqnarray*}}
\newcommand{\eean}{\end{eqnarray*}}

\newcommand{\R}{\mathbb{R}}
\newcommand{\C}{\mathbb{C}}
\newcommand{\Z}{\mathbb{Z}}
\newcommand{\Hb}{\mathbb{H}}

\newcommand{\rSO}{\mathrm{SO}}
\newcommand{\rO}{\mathrm{O}}
\newcommand{\rd}{\mathrm{d}}
\newcommand{\ri}{\mathrm{i}}
\newcommand{\re}{\mathrm{e}}
\newcommand{\rU}{\mathrm{U}}
\newcommand{\rSU}{\mathrm{SU}}
\newcommand{\diag}{\mathrm{diag}}
\newcommand{\rS}{\mathrm{S}}
\newcommand{\rGL}{\mathrm{GL}}
\newcommand{\rPGL}{\mathrm{PGL}}
\newcommand{\rPO}{\mathrm{PO}}
\newcommand{\rSL}{\mathrm{SL}}
\newcommand{\rOL}{\mathrm{OL}}
\newcommand{\rL}{\mathrm{L}}
\newcommand{\rAd}{\mathrm{Ad}}
\newcommand{\rVer}{\mathrm{Ver}}
\newcommand{\rHor}{\mathrm{Hor}}
\newcommand{\rrank}{\mathrm{rank}}
\newcommand{\rdim}{\mathrm{dim}}
\newcommand{\rker}{\mathrm{ker}}
\newcommand{\rspan}{\mathrm{span}}

\newcommand{\cF}{\mathcal{F}}
\newcommand{\cH}{\mathcal{H}}
\newcommand{\cL}{\mathcal{L}}
\newcommand{\cD}{\mathcal{D}}
\newcommand{\cC}{\mathcal{C}}
\newcommand{\cP}{\mathcal{P}}
\newcommand{\cM}{\mathcal{M}}
\newcommand{\cN}{\mathcal{N}}
\newcommand{\cU}{\mathcal{U}}
\newcommand{\cV}{\mathcal{V}}

\newcommand{\e}{\epsilon}

\newcommand{\fso}{\mathfrak{so}}
\newcommand{\fk}{\mathfrak{k}}
\newcommand{\fp}{\mathfrak{p}}
\newcommand{\fgl}{\mathfrak{gl}}

\newcommand{\tA}{\tilde{A}}
\newcommand{\tD}{\tilde{D}}
\newcommand{\te}{\tilde{\eta}}
\newcommand{\tF}{\tilde{F}}
\newcommand{\tN}{\tilde{\nabla}}
\newcommand{\tpar}{\tilde{\partial}}
\newcommand{\tR}{\tilde{R}}

\newcommand{\rdt}{\tilde{\mathrm{d}}}

\begin{flushright}
IFIC/09-39\\
DISTA/PHYS-020/09\\
hep-th/09075583
\end{flushright}
\vspace{.5cm}

  \centerline{\LARGE \bf  Torsion formulation of gravity.}

\vskip 1.5cm

\centerline{
M. A. Lled\'o$^{\sharp\natural}$ and L. Sommovigo$^{\sharp\flat}$ }
 \vskip
1.5cm

\centerline{\it $^\sharp$
  Departament de F\'{\i}sica Te\`{o}rica, Universitat de Val\`{e}ncia,  and }
\centerline{\it IFIC (Centro mixto CSIC-UVEG).}
 \centerline{\small\it C/Dr.
Moliner, 50, E-46100 Burjassot (Val\`{e}ncia), Spain.}

\bigskip

\centerline{\it $^\natural$
 Fundaci\'{o} General Universitat de Val\`{e}ncia.}

\bigskip

\centerline{\it $^\flat$ Dipartimento di Scienze e Tecnologie Avanzate and}
\centerline{\it INFN Gruppo collegato di Alessandria,}
\centerline{\it Universit\`{a} del Piemonte Orientale,}
\centerline{\it Via Teresa Michel 11, 15121 Alessandria, Italy}

\bigskip

 \centerline{{\footnotesize e-mail: \; Maria.Lledo@ific.uv.es,\;
Luca.Sommovigo@mfn.unipmn.it
 }}


\vskip 2cm

\begin{abstract}
We make it precise what it means to have a connection with torsion as solution of the Einstein equations. While locally the theory remains the same, the new formulation allows for topologies that would have been excluded in the standard formulation of gravity. In this formulation it is possible to couple arbitrary torsion to gauge fields without breaking the gauge invariance. \end{abstract}

 \vfill\eject

\section{Introduction}

In Einstein's theory of gravity, the quantity that represents the gravitational field is the pseudoriemannian metric. This comes from the equivalence principle, so the physics of gravity is formulated naturally in a geometric language. Mathematically, (pseudo) Riemannian geometry is based on the existence of a unique, torsionless connection compatible with the metric, the Levi--Civita connection. Although the field equations of gravity are formulated in local coordinate patches, different solutions are  `reconstructions' of the Riemannian manifold  by gluing appropriately different patches, then giving the possibility of vacua that have different topology.

Nevertheless, it is not mandatory to use the Levi--Civita connection, since there are many connections that are compatible with the metric. They have torsion, and it is the existence and uniqueness result for the Levi--Civita connection that makes it the standard choice. There have been many attempts to introduce torsion in gravity. Usually, torsion in the connection is interpreted as new degrees of freedom, new fields with essentially different physical content. One has
$$\Omega=\Gamma+\Delta,$$
where $\Omega$ is an arbitrary (metric) connection, $\Gamma$ is the Levi--Civita connection and $\Delta$ is a tensor that determines univocally the torsion of $\Omega$ (see Section \ref{sec:TR}). Reference \cite{sh} is a very good review article on possible treatments and applications of gravity theories with torsion.

The point of view that we take here is different. We assume that on the manifold there is a background  connection $\omega$ without torsion, which is the Levi--Civita connection of a particular (background) metric. The word `background' is used here in a slightly different way than usual, since this metric does not need to be a solution of Einstein's equations (although it may be convenient to choose it in that way), not even to first order in some perturbation theory. One has to think on it as a  a reference metric, arbitrarily chosen,  and the only assumption that we are making is that the manifold admits a pseudoriemannian metric\footnote{This condition is topological, and it will be explained in detail in Theorem \ref{isomorphicreduction}.}.  The connection that we see as physical is, say, $\Omega$, and the difference between them is a tensor $\Delta=\Omega-\omega$  that, as said before,  determines univocally the torsion. The torsion, then, can be used as the field of gravity, and that is what we call the {\it torsion representation of gravity}.

In the last paragraph our considerations where essentially local. Coming back to the `reconstruction' of the global solution, it is surprising that in the new formulation different gluings are allowed, and topologies that could not be considered in the classical approach appear here as possible vacua. As a particularly interesting example we have the {\it twisted torus} \cite{cs}, used in the literature as a  compactification manifold of supergravity \cite{kstt}. As we will see in Example \ref{twistedtorus}, the interpretation in this formalism is extremely easy and it could provide a simple way of proving that these compactifications are indeed spontaneous \cite{he}.

The other advantage that the formulation presents is that it solves naturally the problem of coupling spacetime torsion to gauge fields \cite{hrrs}. In the classical interpretation the  gauge transformations had to be modified, and a restriction to a particular type of torsion (trace-type torsion) had to be imposed. In the torsion formulation this is not anymore a problem, and the interaction with gauge or matter fields is straightforward.

The paper is organized as follows:

In Section \ref{trep} we justify mathematically the formalism and define the variables of the torsion representation of gravity.

In Section \ref{sec:TR} we write the Cartan--Einstein lagrangian and compute the field equations. Then we proceed to illustrate the appearance of new topologies with two relevant examples.
The first example that we consider is the twisted 3-torus, related locally to Euclidean space by the torsion formulation.
The second example is extracted from an old paper \cite{hr} in which torsion was introduced to modify a gravity solution in order to model a gravitational vortex. This is quite close to our finding that the torsion formulation can be glued to new topologies, although a detailed analysis will show differences of interpretation.

In Section \ref{interacctions}, we show how to couple gravity to gauge and scalar fields  in the torsion formulation, and finally we state our conclusions with prospectives for future work.

\section{Gravitational field in the  torsion representation \label{trep}}

The idea underlying what we call the {\it torsion representation} of the gravitational field is that the same  geometric information can be carried by a connection with torsion or by one without torsion. In this section we are  interested in a global description, so we will always specify the open set in which we are working, and then we will take into account the gluing of the quantities in the intersection of open sets.

In this paper we will follow closely the notation of Reference \cite{kn}.

Let $\cM$ be a manifold of dimension $n$ and let $\begin{CD}\rL(\cM)@>\pi>>\cM\end{CD}$ be its {\it bundle of frames} or {\it frame bundle}. An element of the frame bundle is an arbitrary basis of the tangent space of a point $x\in\cM$.

Let
$\{\cU_{(i)}\}_{i\in I}$ be an open cover of $\cM$. We will indicate the quantities defined on the open set $\cU_{(i)}$ with the subindex
`$(i)$'. Let $\{{V}_{(i)}\}_{i\in
I}$ be local frames  on  each $\cU_{(i)}$ providing a  trivialization
of the frame bundle, so
$$V_{(i)}=\{V_{(i)\bar{a}}(x),\; \bar{a}=1,\dots n,\, \forall x\in \cU_{(i)}\},$$
where $V_{(i)\bar{a}}(x)$ are vectors forming a basis  of the tangent space to the point $x\in \cM$, smoothly depending on $x$.
The $\rGL(n,\R)$-valued transition functions $a_{(ij)}(x)$ are
 given by
\begin{equation}{V}_{(j)}={V}_{(i)}a_{(ij)},\label{transition}\end{equation} (there is no summation on the indices $(i), (j)$) and in the intersection of three open sets they satisfy  the
cocycle condition
$$a_{(ij)}a_{(jk)}=a_{(ik)},\qquad a_{(ij)}^{-1}=a_{(ji)}.$$
The {\it vielbein}  is the dual frame to ${V}_{(i)}$, a basis of one-forms that we will denote  as  $V^{(i)}$, with
\begin{equation}V^{(i)\bar{a}}V_{(i)\bar{b}}=\delta^{\bar{a}}_{\bar{b}}.\label{vielbein}\end{equation}
Let $\{x^\mu, \, \mu =1,\dots n\}$ be a coordinate system in $\cU_{(i)}$. The the frame ${V}_{(i)}$ and the vielbein are
\bean
&&{V}_{(i)}=\{{V}_{(i)\bar a}^\mu\,\partial_\mu,\;\; \;\bar a=1,\dots ,n,\; \mu=1,\dots ,n\},\\
&&{V}^{(i)}=\{{V}^{(i)\bar{a}}_{\mu}\,\rd x^\mu,\; \;\bar{a}=1,\dots ,n,\; \mu=1,\dots ,n\}.
\eean

We consider a connection $\Omega$ on $T\cM$. Associated to the trivialization $\{{V}_{(i)}\}_{i\in
I}$ of the frame bundle, the connection is given by a set of $\fgl(n,\R)$-valued 1-forms $ \{\Omega_{(i)}\}_{i\in
I}$, one on each open set $\cU_{(i)}$,  satisfying the gluing condition
\be \Omega_{(j)}=a_{(ij)}^{-1} \Omega_{(i)}a_{(ij)}-a_{(ij)}^{-1}\rd
a_{(ij)}.\label{gluingcondition}\ee
The connection $\Omega$ is then specified by $\{\Omega_{(i)}, {V}_{(i)}\}_{i\in
I}$.
The curvature is the $\fgl(n,\R)$-valued 2-form
\begin{equation}R_{(i)}=\rd\Omega_{(i)}+\Omega_{(i)}\wedge\Omega_{(i)},\label{curvature}\end{equation} with a tensorial gluing condition,
$$R_{(j)}=a_{(ij)}^{-1}R_{(i)}a_{(ij)},$$
and satisfying the {\it first Bianchi identity}
\begin{equation} \rd R_{(i)} +\Omega_{(i)}\wedge  R_{(i)}=0.\label{1BI}\end{equation}

The torsion 2-form is given in each open set by
$$T_{(i)}=\rd  V^{(i)}+  \Omega_{(i)}\wedge  V^{(i)},$$  and the  gluing in different open sets is
$$T_{(j)}=a_{(ij)}^{-1}T_{(i)}.$$ Together with the curvature, it satisfies the {\it second Bianchi identity}
\begin{equation}\rd T_{(i)} +\Omega_{(i)}\wedge  T_{(i)}=  R_{(i)}\wedge V_{(i)}.\label{2BI}\end{equation}

We
want now to decompose the equation $T_{(i)}= 0$ in an unusual way\footnote{This type of decomposition was suggested to us by L. Andrianopoli. In fact, it has been used before in particular cases, see for example \cite{alt}.}.
We set
\begin{equation}V^{(i)}=M^{(i)}E^{(i)},\qquad {V^{(i)}}^{\bar{a}} = M^{(i)}{}^{\bar{a}}{}_{b} {E^{(i)}}^{b},\label{changesection}
\end{equation} where $M^{(i)}$  is a
$\rGL(n,\R)$-valued function on  $\cU_{(i)}$. We denote also $ (M^{(i)})^{-1}\equiv  M_{(i)}$. Then, $\{E_{(i)a}, \; a=1,\dots, n\}$ are local sections defining another trivialization of the frame bundle.
 They satisfy
\begin{equation}{E}_{(j)}={E}_{(i)}b_{(ij)},\qquad
b_{(ij)}=M^{(i)}a_{(ij)}M_{(j)}.\label{transort}\end{equation} The group valued functions $b_{(ij)}$ satisfy also
 the cocycle condition
$$b_{(ij)}b_{(jk)}=b_{(ik)},$$ so they are also transitions functions
of the frame bundle.

 It is convenient to use indices `$a$' for tensors referred to this new frame and $\bar a$ indices for tensors referred to the former. We have
\begin{equation} T_{(i)}=\rd M^{(i)}\wedge
E^{(i)}+M^{(i)}\rd E^{(i)} + \Omega_{(i)} \wedge
M^{(i)}E^{(i)}=0.\label{decomposition}\end{equation} Notice that we are not changing $\Omega_{(i)}$ as in (\ref{gluingcondition}),
$$\Omega'_{(i)}=M^{(i)}\Omega_{(i)}M_{(i)}-M^{(i)}\rd M_{(i)}.$$ If we were doing so, we will obtain the torsion in the new trivialization $\{{E}_{(i)}\}_{i\in
I}$, $$ {T'}_{(i)}=M^{(i)}T_{(i)}=\rd E^{(i)}
+{\Omega'}_{(i)}\wedge E^{(i)}=0,$$ which of course is also
zero. 
Instead, we transform  $\Omega_{(i)}$ as
$$
 \tilde \Omega_{(i)}=M_{(i)}\Omega_{(i)} M^{(i)},$$ or in components
$$\tilde\Omega_{(i)}{}^{a}{}_{b}=M_{(i)}^{a}{}_{\bar{b}}\,\Omega_{(i)}^{\bar{b}}{}_{\bar{c}} \,M^{(i)}{}^{\bar{c}}{}_{b}.$$
Rearranging
the terms in (\ref{decomposition}) we obtain
\begin{equation} \rd E^{(i)}+\tilde \Omega_{(i)}\wedge E^{(i)}=\rd M_{(i)} \cdot M^{(i)}\wedge E^{(i)}.\label{splittorsion}\end{equation}
We want to interpret  the term, $\rd
E^{(i)}+\tilde \Omega_{(i)}\wedge E^{(i)}$, as the torsion of a new
connection, let us call it  $\tilde \Omega$. The new
connection is $(\tilde \Omega_{(i)},E_{(i)})$ refers to the frame ${E}_{(i)}$.

But we still  have to check that $(\tilde\Omega_{(i)}, {E}_{(i)})$ satisfy the gluing condition.  $\tilde \Omega$ is well defined if in the intersection between two charts if and only if
\begin{equation}\tilde\Omega_{(j)}=b_{(ij)}^{-1}\tilde\Omega_{(i)}b_{(ij)}-b_{(ij)}^{-1}\rd
b_{(ij)}.\label{gluingconnection}\end{equation} By assumption,  the equation that is satisfied is (\ref{gluingcondition}). It is easy to show that (\ref{gluingconnection}) is satisfied if and only if
$$a_{(ji)}(M^{(i)}\rd M_{(i)})a_{(ij)}=(M^{(j)}\rd M_{(j)}).$$ If we denote $M_{(i)}=\exp A_{(i)}$, then the condition is written as
\be a_{(ji)}\rd A^{(i)}a_{(ij)}=\rd A^{(j)}.\label{gluing}\ee
Given the  $a_{(ji)}$ for all pairs $(i,j)$, (\ref{gluing}) imposes a condition on the choice of $M^{(i)}$ and  $M^{(j)}$ for all $(i)$ and $(j)$. But assuming that we find a solution for each pair $(i, j)$, the solution is consistent, since the cocycle condition assures that if
  $$a_{(ji)}X_{(i)}a_{(ij)}=X_{(j)}, \qquad a_{(kj)}X_{(j)}a_{(jk)}=X_{(k)},$$ then
  $$a_{(ki)}X_{(i)}a_{(ik)}=X_{(k)}.$$
  The condition (\ref{gluing}) implies that  the one forms $\rd A^{(i)}$ define a global section of a bundle associated to $\rL(\cM)$ through the adjoint representation.

  \begin{example}{}\hfill

  \begin{itemize}

  \item If $M^{(i)}=\re^{\phi_i}\id$, then (\ref{gluing}) is trivially satisfied for arbitrary transition functions provided $$\phi_{(i)}-\phi_{(j)}=c_{(ij)}=\mathrm{constant}.$$

  \item If the transition functions are of the type $$a=\begin{pmatrix}\alpha&0\\0&\beta\end{pmatrix},$$ then we can have a more general solution with
  $$M=\begin{pmatrix}\re^{\phi_1}&0\\0&\re^{\phi_2}\end{pmatrix}.$$

  \item If the transition functions are constant (the case of a {\it flat bundle}), then $M$ can be arbitrary.\end{itemize}
\hfill$\blacksquare$
  \end{example}

 Let us now introduce  a pseudoriemannian metric in $\cM$, represented locally by $g_{\mu\nu}$ and let $V^{(i)}$ be orthonormal frames
 $$g_{\mu\nu}=\eta_{pq}V^{(i)\bar p}_\mu V^{(i)\bar q}_\nu=\eta_{\bar p\bar q}a_{(ij)}{}^{\bar p}_{\bar r}V^{(j){\bar r}}_\mu b_{(ij)}{}^{\bar q}_{\bar s}V^{(j){\bar s}}_\nu.$$  This means that $a_{(ij)}$ must be valued in $\rO(p,q)$, and that the vielbeins are unique only up to an orthogonal transformation. We denote by $\Omega$ the Levi-Civita connection associated to $g_{\mu\nu}$, locally represented by $\{\Omega_{(i)}, V_{(i)}\}$.

 We assume also that there is another metric with orthonormal frame $E^{(i)}$,
 $$g'_{\mu\nu}=\eta_{pq}E^{(i)p}_\mu E^{(i)q}_\nu=\eta_{pq}b_{(ij)}{}^p_rE^{(j)r}_\mu b_{(ij)}{}^q_sE^{(j)s}_\nu.$$
The matrices $b_{(ij)}$ must be valued in $\rO(p,q)$. Let $\omega$ be the Levi--Civita connection associated to $g'_{\mu\nu}$, which is locally represented by $\{\omega_{(i)}, E_{(i)}\}$.

 Both metrics define different reductions of the frame bundle to orthogonal bundles, $\rOL(\cM)$, $\rOL'(\cM)$. We are going to show that the two reductions are isomorphic.

 \begin{theorem}\label{isomorphicreduction} { \sl Let $g$ and $g'$ be two pseudoriemannian metrics  on $\cM$, generically non isometric but with the same signature $(q,p)$. Let $\rOL(\cM)\subset \rL(\cM)$ be the subbundle of the frame bundle consisting on orthogonal frames with respect to the metric $g$, and similarly $\rOL'(\cM)\subset \rL(\cM)$ the subbundle of orthogonal frames with respect to the metric $g'$. The bundles $\rOL(\cM)$ and $\rOL'(\cM)$, which are principal bundles with structure group $\rO(q,p)$, are isomorphic bundles.}

 { \sl Proof.} Let $U=V^{-1}$ and $U'=V'^{-1}$ be local frames of $\rOL(\cM)$ and  $\rOL'(\cM)$ respectively (see (\ref{vielbein}) (the notation $U$ for the vielbein matrix is introduced here for simplicity). We then have
 $$U^T\eta U=g,\qquad U'^T\eta U'=g'.$$ The metrics $g$ and $g'$ are real symmetric matrices with the same number of positive and negative eigenvalues, so by {\it Sylvester's law of inertia}, they are related by a congruence transformation,
$$g'=S^T gS,$$ where $S$ is an invertible matrix. One then obtains that $U$ and $U'$ can be chosen as
$U'=US.$ The matrix $S$ realizes the local isomorphism. Nevertheless, the matrix  $S$ is not unique, since $\tilde S=OS$, where $O$ is orthogonal with respect to the metric $g$, would also realize the congruence. We then need a criteria to choose uniquely the matrix $S$.

Let us first consider the Riemannian case\footnote{We  thank V. S. Varadarajan for showing us this argument.}. We have then the polar decomposition of $S$,
$S=OP$,  where $O$ is orthogonal (with respect to $g$) and $P$ is a symmetric, positive definite matrix. Since $S$ is invertible the polar decomposition is unique, $P$ is unique and
$$g'=P^TgP,\qquad U'=UP$$ define a global isomorphism of the principal bundles $\rOL(\cM)$ and $\rOL'(\cM)$.

The pseudoriemannian case needs extra work. The polar decomposition is not true anymore with $O\in\rO(q,p)$, but manifolds that admit a pseudoriemannian  metric have some topological restrictions that will allow us to overcome the problem.

First, we note that by the Iwasawa decomposition, a real group $G$ is diffeomorphic to the product of a maximal compact subgroup times an Euclidean space. By Theorem 5.7 in \cite{kn},  any principal bundle over a paracompact basis can be reduced to the maximal compact subgroup. For example, in the case of  the frame bundle, the maximal compact subgroup of $\rGL(n,\R)$ is $\rO(n)$, and each reduction defines a metric on $\cM$. This is another way of proving that there always exists a Riemannian metric over a paracompact manifold.

We have two $\rO(q,p)$-bundles, whose structural groups can be reduced to the maximal compact subgroup, $\rO(p)\times \rO(q)$. We denote these reductions as $\rOL_c(\cM)$ and $\rOL'_c(\cM)$. This has an important implication on the tangent bundle of the manifold. Since the transition functions can be chosen of the form
$$\begin{pmatrix} a&0\\0&b\end{pmatrix},  \qquad a\in\rO(p),\quad b\in\rO(q),$$ then the tangent bundle $T\cM$ of a manifold that admits a pseudoriemannian metric admits a global splitting in two transversal distributions of dimensions $q$ and $p$ respectively,
$$T\cM=T\cM_p\oplus T\cM_q.$$
In fact, this is a sufficient condition for the existence of a pseudoriemannian metric of signature $(p,q)$.

Let $U$ and $U'$ local vielbeins of $\rOL_c(\cM)$ and  $\rOL'_c(\cM)$ respectively (see (\ref{vielbein})).   Due to the global splitting of the tangent bundle we can in fact choose in every open set
$$U=\begin{pmatrix} u&0\\0&v\end{pmatrix}, \qquad U'=\begin{pmatrix} u'&0\\0&v'\end{pmatrix},$$ and the transition functions will not change this form. We can in fact work with the subbundles $T\cM_p$ and $T\cM_q$ and apply the polar decomposition argument in each of them.

We have then proven that the reductions of the frame bundle associated to two pseudoriemannian metrics, not necessarily isometric, are isomorphic bundles. \hfill$\blacksquare$

 \end{theorem}

 If the condition (\ref{gluing}) is satisfied, then by the procedure described  in Eq. (\ref{decomposition}) we can associate a connection with torsion  obtained by gluing $\{\tilde\Omega_{(i)}, E_{(i)}\}$. This is an alternative to the Levi-Civita connection to describe the degrees of freedom of the gravitational field. Physically, $g'_{\mu\nu}$ (or $E_{(i)}$) and $\omega$ are fixed, reference quantities, and the physical variables are $M$ and the torsion $T$, or equivalently
 $$\Delta=\Omega-\omega, $$ linked by (\ref{decomposition}). It
 is what we call the torsion representation of the gravitational field.

The equations of motion are only local, and the obstruction to the gluing  is already indicating that the topologies of the global solution may have to be different.

\section{Cartan--Einstein action in the torsion representation}

\label{sec:TR}

We want now to write the action principle for the gravitational field in the torsion representation. The field equations are local equations. They are written in an open set of the manifold, so we will drop the index $(i)$. We will also suppress the symbol ``$\wedge$'' for the
wedge product of $n$-forms, as it is customary in physics
notation. The meaning should be clear from the context.

We are using two different frames, $V$ and $E$. As before,
the indices of tensors referred to the frame $V$ will be written as
`$\bar a$', while the indices referred to the frame $E$ will be `$a$',
so, for example,
$$V^{\bar a} = V^{\bar a}_\mu \, \rd x^\mu, \qquad  E^a=E_\mu^a \, \rd
x^\mu,\qquad V^{\bar a}= {M^{-1}}^{\bar a}{}_a E^a.$$

Let us consider
the Cartan--Einstein Lagrangian in $D=4$\footnote{We consider only
  $D=4$ for concreteness and not to load the notation, but it is clear
  that the same argument can be applied in any dimension.} in first
order formalism\footnote{Introduced in the context of the so called geometric or rheonomic approach to Supergravity \cite{cdf}.} (with independent variables $V$ and $\Omega$)
\be\cL_G = - \frac{1}{4} \e_{\bar a\bar b\bar c\bar d} R^{\bar a\bar b}
V^{\bar c} V^{\bar d}\bigl(=- \frac{1}{4} \e_{\bar a\bar b\bar c\bar
  d}   R^{\bar a\bar b}\wedge V^{\bar c}\wedge V^{\bar
  d}\bigr),\label{eclagrangian}\ee
where $R^{\bar a\bar b}=R^{\bar a}{}_{\bar{c}}\eta^{\bar{c}\bar{ b}}$
is the Riemann curvature  two-form as in Eq.(\ref{curvature}).
Varying the Lagrangian with respect to $V$ we obtain the Einstein equations,
\begin{equation}R^{\bar a\bar b}{}_{\bar{c}\bar{ b}} - \frac{1}{2}
  \delta^{\bar a}_{\bar{c}} R=0, \label{Einstein}\end{equation}
with the curvature two-form expressed in the vielbein basis as
$$R^{\bar a\bar b}= R^{\bar a\bar b}{}_{\bar{c}\bar{ d}}  V^{\bar c} V^{\bar d},
\,\hbox{ and  } R=R^{\bar a\bar b}{}_{\bar a\bar b}.$$
Variation with respect to $\Omega$ gives the torsionless condition
\begin{equation} {T(V,\Omega)}^{\bar a} = \rd V^{\bar a} +
  \Omega^{\bar a}{}_{\bar b} V^{\bar b} = 0. \label{torsion}
\end{equation}

\medskip

We perform now the  local change of variables. We assume that $E$
is a fixed, arbitrary vielbein. We want to refer all the quantities to the frame
$E$. The vielbein $E$ is associated to a metric $g'_{\mu\nu} =
\eta_{ab} E^a_\mu E^b_\nu$. But the frame $V$ is not
orthonormal with respect to this metric:
$$V_{\bar a}^\mu V_{\bar b}^\nu g'_{\mu\nu}=\eta_{ab}M^a{}_{\bar a}M^b{}_{\bar b}=h_{\bar a\bar b}\neq
\eta_{\bar a\bar b}.$$  Similarly
\be E_a^\mu E_b^\nu g_{\mu\nu}=\eta_{\bar a\bar b}(M^{-1})_a{}^{\bar a}(M^{-1})_b{}^{\bar b}=h_{ab}\neq
\eta_{ab},\label{fibermetric}\ee with $g_{\mu\nu} =
\eta_{ab} V^a_\mu V^b_\nu$.

In the original
Lagrangian (\ref{eclagrangian}), $\Omega$ is the one-form representing
the Levi--Civita connection with respect to $V$. In
the $E$ basis,  $$\tilde\Omega^a{}_b=M^a{}_{\bar a}\Omega^{\bar a}{}_{\bar
  b}{M^{-1}}^{\bar b}{}_b,$$ which together with $E$ define a
 connection with torsion
\begin{equation}
 {T(E,\tilde\Omega)}^{a} = D E^{a} =\rd E^{a} + \tilde\Omega^{a}{}_{b} E^{b}.
\label{deltat}\end{equation}

Let $\omega$ be the torsionless  connection associated to $E$, so
\begin{equation}
\mathcal{D} E^{a} \equiv \rd E^{a} + \omega^{a}{}_{b} E^{b} = 0. \label{omeganot}
\end{equation}
We define the tensor $\Delta$ as
$$\tilde\Omega^{a}{}_{b}=\omega^{a}{}_{b}+\Delta^a{}_b.$$ $\Delta$ carries all the information about the connection $\Omega$. Moreover,
$$T(E,\tilde\Omega)^{a}=\Delta^a{}_bE^b,$$ where we have used (\ref{omeganot}).
The map sending  $\Delta_{b}{}^{a}{}_{c}$ to $T^{a}{}_{{b}{c}} = \Delta_{[{b}}{}^{a}{}{}_{{c}]}$ is one to one (remember that $\Delta_{{b}{a}{c}}$ is antisymmetric in the indices $a, c$). Its inverse is given by
\begin{align*}
\Delta_{{b}{a}{c}} = \Delta_{b}{}^{e}{}_{c} h_{{e}{a}}, \qquad T_{{a}{b}{c}} = h_{{a}{e}} T^{e}{}_{{b}{c}} \\
\Delta_{{b}{a}{c}} = \frac{1}{2} (T_{{a}{b}{c}} - T_{{c}{b}{a}} + T_{{b}{a}{c}}).
\end{align*}

\begin{remark}This fact can be used to show the existence and uniqueness of a torsionless connection compatible with an $\rO(p,q)$-structure \cite{jo}. One just starts from an arbitrary connection $\tilde\Omega$ with torsion $T$ and uses the bijectivity of the map to give $\Delta$, and in turn $\omega$, which is torsionless. One can then say  that any  $\rO(p,q)$-structure is {\it torsionless }.\hfill$\blacksquare$\end{remark}

The connection $\omega$ and the vielbein $E$ are fixed (for instance they can be the ones of flat Minkowski spacetime if the topology allows for it). Then all the dynamics of the system lies in $M$ and $\Delta$ (or $T$). We stress that the equations are exact, and no approximation has been done.

We denote by $\mathcal{R}^a{}_b$  the Riemann tensor of the connection $\omega$, so it does not depend on the dynamical variables. The symbol $\mathcal{D}$ stands for the covariant derivative with respect to $\omega$.
Explicitly,
\bean\mathcal{R}^a{}_b &=& \rd \omega^a{}_b + \omega^{a}{}_{c}
\omega^c{}_b, \qquad \mathcal{R}^a{}_b = \mathcal{R}^a{}_{bcd} E^{c}
E^{d},\\
\cD \Delta^a{}_b &=& \rd \Delta^a{}_b + \omega^a{}_c\,
\Delta^c{}_b + \Delta^a{}_c\,
\omega^c{}_b .\eean
Also, $h^{ab}$ denotes the inverse of $h_{ab}$ defined in  (\ref{fibermetric}).
In terms of the new variables, the Einstein--Cartan Lagrangian (\ref{eclagrangian}) becomes
\bea\cL_G& = &\frac{1}{4 \det{M}}  \bigl[\mathcal{R}^a{}_b + \mathcal{D}
\Delta^a{}_b + \Delta^{a}{}_{c} \Delta^c{}_b -(\omega^a{}_c+\Delta^a{}_c)( M^{-1}\rd M)^c_b-
\nonumber\\
&&
\qquad\qquad -(M^{-1}\rd M )^a_c(\omega^c{}_b+\Delta^c{}_b)\bigr]h^{eb}\epsilon_{aecd}E^cE^d\label{gravitylagrangian}\eea

For simplicity, we set now $\omega=0$ and $E^a_\mu=\delta^a_\mu$, so the background is Minkowski space. In terms of the new variables, the equations of motion read
\bea
&&\Delta_{[a}{}^c{}_{b]} = - F^c{}_{[ab]} \label{grav1}\\
&&\partial_c \Delta_a{}^c{}_b - \partial_b \Delta_c{}^c{}_a +
\Delta_c{}^c{}_d \Delta_b{}^d{}_a - \Delta_b{}^c{}_d \Delta_c{}^d{}_a -
\nonumber \\
&&- F^c{}_{dc} \Delta_b{}^d{}_a - F^d{}_{ab} \Delta_c{}^c{}_d + F^c{}_{db}
\Delta_c{}^d{}_a + F^d{}_{ac} \Delta_b{}^c{}_d = 0\label{grav2}
\eea
where the square brackets mean antisymmetrization in the indices,  and we have defined
$$F^p{}_{qr}=\partial_rM^p{}_{\bar p}(M^{-1})^{\bar p}{}_q.$$

Equation (\ref{grav1}) is a constraint between $\Delta$ and $M$. It is equivalent to  (\ref{torsion}), and its geometrical meaning has been made clear previously.
Equation (\ref{grav2}) is  equivalent to the Einstein's
equations (\ref{Einstein}).

\begin{remark}\end{remark} The quantities $\omega$ and $E$ are related by the zero torsion equation (\ref{omeganot}), but they do not need to satisfy Einstein's equations (\ref{Einstein}). This  is then a matter of convenience.
\hfill$\blacksquare$

\medskip

We want now to show how Euclidean  spacetime can be locally but not globally, related by this change of variables to manifolds with topology that do not admit the flat, Euclidean metric. This is the spirit of Examples \ref{twistedtorus} and \ref{meisner}. The setting   is different from the setting in (\ref{grav1},\ref{grav2}), since the background is not Euclidean space. In fact, it would not be too difficult to add the suppressed terms to the equations, but this will be not necessary for our analysis.

\begin{example}{\it The twisted 3-torus.}\label{twistedtorus}

Let us start with the Euclidean space $\R^3$ with metric
\begin{equation}
\rd s^2 = \rd x^2 + \rd y^2 + \rd z^2.
\end{equation}
The simplest vielbein is
\be V^a = \begin{pmatrix}  \rd x\\ \rd y\\ \rd z \end{pmatrix},\label{coordv}\ee and the torsionless connection is just $\Omega = 0$. This is trivially a solution of the Einstein equations (\ref{Einstein}).

Now we consider a different (`twisted') vielbein on $\R^3$
\be E = \begin{pmatrix} \rd x\\\rd y\\\rd z + Nx \rd y\end{pmatrix}.\label{newv}\ee $E$ and $V$ are related by the matrix
\begin{equation}
M=
\begin{pmatrix}
1&0&0\\
0&1&0\\
0&Nx&1
\end{pmatrix},\qquad V=(M^{-1}) E. \label{emme}
\end{equation}
The pair $(\tilde\Omega=0, E)$ define a connection with torsion,
$$T(E,\Omega)^a=\rd E^a=\delta^a_ 3N E^1E^2.$$
The `twisted' vielbein has associated a torsionless connection,
\be\omega^1{}_2=-\frac N 2 E^3, \qquad \omega^1{}_3 = - \frac N 2 E^2, \qquad \omega^2{}_3=\frac N 2 E^1,\ee
 and the rest zero. Then
 $\Delta = \tilde\Omega - \omega = - \omega$ and $M$ are solutions of (\ref{grav1},\ref{grav2}), equivalent to the flat space solution $(V, \Omega)$ of the Einstein equations.

 We can also define the  `twisted' metric on $\R^3$ using the `twisted'  vielbein
\be \rd {s'}^2=\rd x^2+\rd y^2+(\rd z +Nx\rd y)^2.\label{newm}\ee
This metric is not a solution of the Einstein equations

 Notice that we could have chosen any vielbein in $\R^3$ and perform the same trick. We would then obtain different ways of describing the flat space $\R^3$, solution  of the Einstein equations. But the vielbein $E$ can be defined on a compact space constructed by identifying points in $\R^3$. A 3-torus, with the flat metric, is obtained
  by identifying
$$x=x+a,\qquad y=y+b,\qquad z=z+c.$$
If one modifies these conditions in the form
$$x=x+a, \qquad y=y+b,\qquad z= z-Nay+c,$$ one finds that the  `twisted' vielbein (\ref{newv}) is globally defined. The resulting compact manifold is then parallelizable. It is in fact the {\it twisted 3-torus} \cite{cs}. This manifold is a non trivial $T^2$-bundle over the base $S^1$ (described by  the cyclic coordinate $y$). It does not admit a flat metric (see for example Ref. \cite{wo} for a classification of flat 3-manifolds), but it has a non abelian three dimensional group of isometries (the Heisenberg group).

The coordinate vielbein (\ref{coordv}) still exists locally, although it does not extend to a global frame of the manifold. When a coordinate vielbein is global one says that the manifold is {\it integrably parallelizable}, and this is a very stringent condition. In fact, only $\R^n$, $n$-tori or products of them are integrably parallelizable \cite{con}.

We see in this way that the torsion description of the gravitational field allows for a solution to the Einstein equations which is locally equivalent to flat space but that is topologically different. More generally, what we have shown is that both formulations are equivalent only locally, and that the torsion formulation allows for vacua with topologies that are excluded from the standard formulation.

The twisted torus has been used in the literature to produce different compactifications of supergravity and superstring theory \cite{kstt}.  Needless to say, in the compactification of any theory, the size of the compact manifold is very small compared to  testable distances, so the global and topological effects are crucial.

\hfill$\blacksquare$

\end{example}

\begin{example}{\it Gravitational Meisner effect.}\label{meisner}

This appeared in Ref. \cite{hr}, and it compares the presence of torsion in a certain region of spacetime with the magnetic vortex lines that appear in a semiconductor. We rewrite here the gravitational version and interpret it in terms of the torsion formulation of gravity.

One  starts  with flat, Euclidean space $\R^4$ with coordinates $\{x^\mu,\; \mu=0,\dots, 3\}$. We may see it as the algebra of quaternions $\mathbb{H}\approx\R^4$,
$$q=x^0\id+x^1\sigma_1+x^2\sigma_2+x^3\sigma_3,$$ with the Hamilton product,  given by
$$\sigma_i^2=-1,\qquad \sigma_i\sigma_j=-\delta_{ij}\id+\epsilon_{ijk}\sigma^k,\qquad i,j,k=1,2,3$$ so
\bean q&=&x^0\id+x^1\sigma_1+x^2\sigma_2+x^3\sigma_3,\qquad q'\;=\;y^0\id+y^1\sigma_1+y^2\sigma_2+y^3\sigma_3\\&&\\q\cdot q'&=&(x^0y^0-x^1y^1-x^2y^2-x^3y^3)\id+
(x^1y^0+x^0y^1-x^3y^2-x^2y^3)\sigma_1+\\&+&(x^2y^0+x^3y^1+x^0y^2-x^1y^3)\sigma_2+
(x^3y^0-x^2y^1+x^1y^2+x^0y^3)\sigma_3\eean
The subset $S^3\subset \Hb$
$$S^3=\{q\in\mathbb{H}\, |\, \parallel\! q\!\parallel^2=\delta_{\mu\nu}x^\mu x^\nu=1 \},$$ is indeed the group $\rSU(2)$, with the group law being the Hamilton product. The group acts on  $\R^3=\rspan\{\sigma_1,\sigma_2,\sigma_3\}$ as rotations
$$\overrightarrow{v}'=q\overrightarrow{v}q^{-1},\qquad \overrightarrow{v}\in \R^3,$$ where
$$q^{-1}=\frac 1{r^2}\bigl(x^0\id-x^1\sigma_1-x^2\sigma_2-x^3\sigma_3\bigr),$$ 
 with $r=\parallel\!\! q\!\!\parallel$. One can obtain the Maurer--Cartan forms for $\rSU(2)$ by computing them first in $\Hb$:
\bean q^{-1}\rd q&=&\frac 1{r}\bigl(E^0\id+E^1\sigma_1+E^2\sigma_2+E^3\sigma_3\bigr),\quad \hbox{with}\\
E^0&=&\frac 1{r}\bigl(x^0\rd x^0+x^1\rd x^1+x^2\rd x^2+x^3\rd x^3\bigr)=\rd r,\\
E^1&=&\frac 1{r}\bigl(-x^1\rd x^0+x^0\rd x^1+x^3\rd x^2-x^2\rd x^3\bigr),\\
E^2&=&\frac 1{r}\bigl(-x^2\rd x^0-x^3\rd x^1+x^0\rd x^2+x^1\rd x^3\bigr),\\
E^3&=&\frac 1{r}\bigl(-x^3\rd x^0+x^2\rd x^1-x^1\rd x^2+x^0\rd x^3\bigr).\eean
When restricted to $S^3$, $E^0=0$ and $E^i$, $i=1,2,3$ are the Maurer--Cartan forms of $\rSU(2)$.

$E=\{E^a,a=0,\dots, 3\}$ is a frame that is well defined everywhere except at the origin. Moreover, one can check that
$$\rd s^2= E^0E^0+E^1E^1+E^2E^2+E^3E^3=(\rd x^0)^2+(\rd x^1)^2+(\rd x^2)^2+(\rd x^3)^2,$$ so $E$ is a vielbein for the Euclidean metric on $\R^4-\{0\}$. It is not difficult to check that
\bea \rd E^0&=&0,\nonumber\\ \rd E^1&=&\frac 1{r}E^0\wedge E^1-\frac 2{r}E^2\wedge E^3,\nonumber\\ \rd E^2&=&\frac 1{r}E^0\wedge E^2+\frac 2{r}E^1\wedge E^3,\nonumber\\ \rd E^3&=&\frac 1{r}E^0\wedge E^3-\frac 2{r}E^1\wedge E^2.\label{vhr}
\eea
Then, the Levi--Civita connection $\omega$ given by the torsionless condition
$D E^a=\rd E^a+\omega^a{}_b\wedge E^b=0$ becomes
\be\omega^i{}_k=\frac 1{r}\epsilon^{i\phantom{j}}_{\phantom{i} jk}E^j, \quad \omega^i{}_0=\frac 1{r}E^i,\quad \omega^{ab} = - \omega^{ba}.\label{lchr}\ee

So we have two vielbeins, $E$ and the standard one $V=(\rd x^\mu, \,\mu=0,\dots 3)$ related by a matrix $M$
$$E=MV,\qquad M=\begin{pmatrix}x^0& x^1&x^2&x^3\\-x^1&x^0&x^3&-x^2\\-x^2&-x^3&x^0&x^1\\-x^3&x^2&-x^1&x^0\end{pmatrix}.$$ The Levi--Civita
connections are $\Omega=0$ for $V$ and $\omega$ as defined in (\ref{lchr}), so
$$\Delta =\tilde\Omega- \omega=-\omega$$ as for the case of the twisted torus.
The torsion of the connection defined by $\tilde\Omega $ and $E$ is
$T(E, \Omega)^a=\rd E^a$, which is computed in (\ref{vhr}). Differently from the twisted torus, the metrics defined by $V$ and $E$ are the same.

In Ref. \cite{hr} the authors perform a regularization of $\omega$ by substituting the factor $1/r$ in front of the connection one forms (\ref{lchr}) by a function $\varphi(r^2)$ with appropriate asymptotic behavior. One requires that $\varphi$ is regular when  $r\rightarrow 0$ and that it reproduces $\omega$ in Eq.(\ref{lchr}) when $r\rightarrow \infty$. In this way, the connection can be extended over $r=0$ and also for $r<0$, so one ends up with a space which is topologically $S^3\times \R$.

The connection has torsion, and the vielbein remains singular at $r=0$. The manifold $S^3\times \R$, though, is parallelizable, so there must exist a global frame.

We propose a different procedure,  similar to the one used for the twisted torus, to regularize both, vielbein and connection. The topology will be $S^3\times \R$ and the connection will be the Levi--Civita connection associated to the global frame. The metric is not flat, so it is not solution of the Einstein equations. But, with the torsion formulation of gravity, it can be locally related to the flat metric through a connection with torsion.

Hyperspherical coordinates on $\R^4$ are  useful,
\bean x^0&=&r\cos \psi,\\
x^1&=& r\cos\phi\sin\theta\sin\psi,\\
x^2&=& r\sin\phi\sin\theta\sin\psi,\\
x_3&=& r\cos\theta\sin\psi,\qquad \quad\qquad \psi,\theta\in[0,\pi[,\quad \phi\in[0,2\pi[.\eean Using the program Mathematica\footnote{Wolfram
Research, Inc., Mathematica, Version 5.1, Champaign, IL (2004).} we obtained for the vielbein
\bean E_0&=&\rd r,\\
E_1&=& r\bigl[(\cos\phi\sin\theta)\rd \psi+\sin\psi\sin\theta (-\cos\psi\sin\phi+\sin\psi\cos\phi\cos\theta)\rd\phi+\\&&
\sin\psi(\cos\theta\cos\psi\cos\phi+\sin\phi\sin\psi)\rd\theta\bigr]\equiv r a_1\\
E_2&=& r\bigl[(\sin\phi\sin\theta)\rd \psi+\sin\psi\sin\theta(\cos\psi\cos\phi+\sin\psi\sin\phi\cos\theta)\rd\phi+\\&&
\sin\psi(\cos\psi\sin\phi\cos\theta-\cos\phi\sin\psi)\rd\theta\bigr]\equiv r a_2\\
E_3&=&r\bigl[\cos\theta\rd\psi-\sin^2\psi\sin^2\theta\rd \phi-\cos\psi\sin\psi\sin\theta\rd\theta\bigr]\equiv r a_3
\eean
The one forms $a^i=E^i/r$ and $\rd r$ form in fact a global frame on $S^3\times \R$. We define the new vielbein as
$$ \hat E^0=f(r)E^0,\qquad E^i=h(r) a^i,$$ where $f(r^2)$ and $h(r^2)$ are strictly positive functions. We impose the following asymptotic conditions
\bean &f(r)\xrightarrow[r\rightarrow 0]{}\mathrm{regular},\qquad &f(r)\xrightarrow[r\rightarrow \pm\infty]{}1,\\&h(r)\xrightarrow[r\rightarrow 0]{}\mathrm{regular},\qquad &h(r)\xrightarrow[r\rightarrow \pm\infty]{}|r|.\eean One possible choice is
\be f(r)= 1,\qquad h(r)=\sqrt{r^2+a^2},\qquad a\in\R.\label{regconnection}\ee If we compute the differentials of the one-forms we get
$$\rd\hat E^0=0,\qquad \rd\hat E^i=A(r)\,\hat E^0\wedge E^i-B(r)\,\epsilon^i{}_{jk}\, \hat E^j\wedge \hat E^k,$$ where
$$A=\frac {h'}{fh},\qquad B=\frac 1 h.$$
The connection is then
$$\hat \omega^i{}_0=A(r)\,\hat E^i,\qquad \hat \omega^i{}_j= B(r)\,\epsilon^i{}_{jk} \hat E^j,$$ and the metric  is
$$\rd{\hat s}^2=f(r)^2\rd r^2+ h(r)^2\bigl[\rd\psi^2+\sin^2\psi(\rd\theta^2+\sin^2\theta\rd \phi)\bigr].$$

For the choice (\ref{regconnection}) we get
$$A=-\frac r{r^2+a^2},\qquad B=\frac 1{\sqrt{r^2+a^2}}.$$

\hfill$\blacksquare$

\end{example}

\section{Interactions in the torsion representation\label{interacctions}}

In this section we compute the interaction of gravity in the torsion formulation with abelian gauge fields and scalar fields. We use also the first order formalism \cite{cdf}.

\subsection{Gauge fields}
\label{gt}

We want to consider now an abelian gauge field\footnote{The generalization to non abelian gauge theories is straightforward, but we prefer to keep  the discourse as simple as possible.} with potential $A_\mu$ coupled to gravity. Then we will perform the change of variables to the torsion formulation and we will obtain the precise form of the interaction of gauge fields with torsion. We will use a first order formalism \cite{cdf}. To the Cartan--Einstein Lagrangian (\ref{eclagrangian}) we  add the following term
\be\cL_V = \frac{1}{4} \left( F f_{\bar a\bar b} V^{\bar a} V^{\bar b} + \frac{1}{24} f^{\bar e\bar g} f_{\bar e\bar g} \e_{\bar a\bar b\bar c\bar d} V^{\bar a} V^{\bar b} V^{\bar c} V^{\bar d} \right)\label{lagrangianvector}\ee
with
$F = \rd A = F_{\bar a\bar b} V^{\bar a} V^{\bar b}$ and $ A=A_{\bar a}V^{\bar a}.$ The symbol  $f_{\bar a\bar b}$ stands for an auxiliary field, antisymmetric in the indices $(\bar a,\bar b)$, and the barred indices   are   lowered and raised with the (pseudo) Euclidean metric $\eta_{\bar a\bar b}$ and its inverse $\eta^{\bar a\bar b}$. In components, we have
$$F=\rd (A_{\bar a}V^{\bar a})=\rd (A_{\bar a})V^{\bar a}+A_{\bar a}\,\rd V^{\bar a}=(D A_{\bar a})V^{\bar a},$$ with
$$(D A_{\bar a})=\rd A_{\bar a}-\Omega^{\bar b}{}_{\bar a}A_{\bar b}.$$ We will use the following notation:
$$\partial_{\bar a}=V_{\bar a}^\mu\partial_\mu,\qquad D_{\bar a}=V_{\bar a}^\mu D_\mu,\qquad \bar \Omega^{\bar c}{}_{\bar a} =\Omega_{\bar b}{}^{\bar c}{}_{\bar a}E^{\bar b},$$ so
$$F=D A_{\bar a}V^{\bar a}=(D_{\bar b} A_{\bar a})V^{\bar b}V^{\bar
  a}.$$ Notice that $\Omega_{\bar b}{}^{\bar c}{}_{\bar a}$ is not
necessarily  symmetric in ($\bar a$, $\bar b$), although the
connection is torsionless, because we are not using a coordinate
frame.

Varying the Lagrangian with respect to the auxiliary field $f_{\bar a
  \bar b}$ one obtains
\be f_{\bar a\bar b}=\frac 12\e_{\bar a\bar b\bar c\bar d}F^{\bar
  c\bar d}={}^*\!F_{\bar a\bar b},\label{feaux}\ee
where the symbol `${}^*$' means the Hodge-star operator on
differential forms. Variation with respect to $A$ then gives
\be Df_{\bar a\bar b}=0, \quad \hbox{equivalent to} \quad
\partial_{[\bar c}\,{}^*\!F_{\bar a\bar b]} - \Omega_{[\bar c}{}^{\bar
  d}{}_{\bar b} \, {}^*\!F_{\bar a] \bar d} - \Omega_{[\bar c}{}^{\bar
d}{}_{\bar a} \,{}^*\!F_{\bar d\,|\bar b]}=0.\label{feA}\ee We have used (\ref{feaux}).
Together with the Bianchi identity
$$D F_{\bar a\bar b} = 0 \quad \hbox{equivalent to} \quad \partial_{[\bar
  c} F_{\bar a \bar b]} - \Omega_{[\bar c}{}^{\bar d}{}_{\bar b}
F_{\bar a] \bar d} - \Omega_{[\bar c}{}^{\bar d}{}_{\bar a} F_{\bar d
  | \bar b]} = 0,$$ they are
the equations of the electromagnetic field in presence of gravity.

Let us now perform the change of variables from $\Omega$ and $ V$ to
$\Delta$ and $M$. It is also convenient to make a change of variables
in the auxiliary field,
$${f}_{ab} ={f}_{\bar c\bar d} (M^{-1})^{\bar c}{}_a (M^{-1})^{\bar
  d}{}_b.$$ Then we have$$ f^{\bar a\bar
  b}f_{\bar a\bar b}=f^{ab}f_{ab}.$$ The Lagrangian becomes
\begin{equation}
\cL_V = \frac{1}{4} \left( F f_{ab} E^a E^b + \frac{1}{24\det M} f^{eh} f_{eh} \e_{abcd} E^a E^b E^c E^d \right), \label{lagvectortorsion}
\end{equation}
where \
\bean F&=&\rd(A_aE^a)=(\rd A_a)E^a+A_a\rd E^a= (\cD A_a)E^a,\\ \cD
A_a&=&\rd A_a- \omega^b{}_a A_ b.\eean Notice that in
(\ref{lagvectortorsion}) the electromagnetic potential appears coupled
only to $M$ and not to $\Delta$. Apart from the definition of
$f^{ab}$, which reflects the fact that we are expressing our vectors
in a frame that is not orthogonal with respect to the original metric,
$M$ appears only through its determinant.  For what regards to the
electromagnetic field, $M$ could just be of the form
$M=e^{\phi}\id$, as assumed in \cite{hrrs}, which gives, trough the
constraint (\ref{splittorsion}) a trace type torsion.  This does not
mean that having a more general $M$ and more general torsion is
inconsistent with the electromagnetic field: it
means that such field only couples to the trace component of the torsion.

Let us compute the field equations. Varying with respect to $f_{ab}$ gives
\be f^{ab}= \det M \, {}^* \! F^{ab},\quad \hbox{or}\quad  f_{ab}= {\det M}^{-1} \, {}^* \! F_{ab},\label{gaugefe1}\ee
which differ from (\ref{feaux}) in the factor $\det M$. Varying with
respect to $A$ we get
\be \cD f_{  a  b}=0, \quad \hbox{equivalent to} \quad
\partial_{[  c} \, f_{  a  b]} - \omega_{[  c}{}^{
  d}{}_{  b} f_{  a]  d} - \omega_{[  c}{}^{  d}{}_{
  a} \, f_{  d\,|  b]}=0.\label{gaugefe2}\ee
Formally this equation is identical to (\ref{feA}), but the presence
of the torsion is encoded in the auxiliary field $f$ through the
matrix $M$.
It is worthy to note that the Lagrangian is gauge invariant by
construction. So we have constructed a Lagrangian of a gauge field in
interaction with spacetime torsion which is fully gauge invariant in
the ordinary sense. It is perhaps instructive to compute the tensor
$F_{ab}$  and see how a gauge transformation affects it. We write it
in terms of $F_{\mu\nu}$, the components of the tensor with respect to a
coordinate frame, that remain invariant (covariant in the non abelian
case) under a gauge transformation.
\bean F_{ab}=E_a^\mu E_b^\nu F_{\mu\nu}&=&\frac 12\bigl(E_a^\mu
E_b^\nu(\partial_\mu A_\nu -\partial_\nu A_\mu)=\frac 12
(\partial_aA_b-\partial_bA_a-\\&&(\partial_a E^\nu_b
)E_\nu^cA_c+(\partial_b E^\nu_a )E_\nu^cA_c\bigr).\eean Making a gauge
transformation
$$A'_a=A_a +\partial_a \lambda,$$ we get
$$F'_{ab}=F_{ab}+\frac 12
\bigl(\partial_a\partial_b\lambda-\partial_b\partial_a\lambda-(\partial_a
E^\nu_b )E_\nu^c\partial_c\lambda+(\partial_b E^\nu_a
)E_\nu^c\partial_c\lambda\bigr).$$ But
$\partial_a=E_a^\mu\partial_\mu$ does not commute with $\partial_b$,
and the commutator  exactly cancels
the last terms in the equation above. So $F'_{ab}=F_{ab}$ and we have
checked explicitly the gauge invariance of the action.

\subsection{Scalar fields}
\label{sec:sf}
Finally we consider the Lagrangian of a scalar field charged under $U(1)$ (the analysis can be easily generalized to $n$ scalar fields). As for the gauge field, we introduce an auxiliary field $\phi_{\bar a}$, and the $\rU(1)$ covariant derivative will be denoted as
$$\nabla \phi = \rd \phi + A \phi = (\partial_{\bar a}\phi+A_{\bar a}\phi)V^{\bar a}.$$ For the auxiliary field, the covariant derivative includes also a gravity part,
$$\nabla_{\bar b}\phi^{\bar a}=\partial_{\bar b}\phi^{\bar a}+A_{\bar b}\phi^{\bar a}+\Omega^{\bar a}{}_{\bar c}\phi^{\bar c}.$$
 Then the lagrangian is
$$\cL_S = - \frac{1}{6} \left( \nabla \phi \, \e_{\bar a\bar b\bar c\bar d}\,\phi^{\bar a} V^{\bar b} V^{\bar c} V^{\bar d} - \frac{1}{8} \phi^{\bar r} \phi_{\bar r} \,\e_{\bar a\bar b\bar c\bar d}\,V^{\bar a} V^{\bar b} V^{\bar c} V^{\bar d} \right),$$
and  the field equations are just
$$\phi_{\bar a}-\nabla_{\bar a} \phi=0,\qquad \nabla_{\bar a}\phi^{\bar a}=0,$$ which eliminating the auxiliary field are
$$\partial_{\bar a}\partial^{\bar a}\phi=0.$$

Let us now perform the change of variables $$V^{\bar a}= {M^{-1}}^{\bar a}{}_a E^a,\qquad \phi^{\bar a}{M^{-1}}^{\bar a}{}_a \phi^a.$$ Also, the covariant derivative will be expressed in the frame $E$:
$$\nabla \phi =  (\partial_{ a}\phi+A_{ a}\phi)E^{ a}.$$ Then, the Lagrangian becomes
\be
\cL_S = - \frac{1}{6\det M} \left( {\nabla} \phi {\phi}^a \e_{abcd} V^b V^c V^d - \frac{1}{8} {\phi}^m {\phi}_m \e_{abcd} V^a V^b V^c V^d \right).\label{scalarlag}
\ee
 It is  convenient make  a redefinition of the auxiliary field, $$\tilde \phi^a=\frac{1}{\det M}{\phi}^a,\qquad \tilde \phi_a={\det M}\,{{\phi}_a},$$ so the field equations are
 $$\det M \,\tilde\phi_{ a}-\nabla_{ a} \phi=0,\qquad \nabla_{ a}\tilde \phi^{ a}=0,$$ which imply
 \be\nabla_a\left(\frac1{\det M}{\nabla^a\phi} \right)=0.\label{scalarfe}\ee

\subsection{Complete Lagrangian and equations of motion}
\label{sec:cl}
Summarizing (\ref{gravitylagrangian}, \ref{lagvectortorsion}, \ref{scalarlag}), the first order Lagrangian for vector and scalar fields coupled to gravity in the presence of torsion can be written as:
\bea\cL& =& \cL_G+\cL_V+\cL_S=\frac{1}{ 4\det{M}}  \Bigl[\bigl(\mathcal{R}^a{}_b + \mathcal{D}
\Delta^a{}_b + \Delta^{a}{}_{c} \Delta^c{}_b +
\nonumber\\
&& -(\omega^a{}_c + \Delta^a{}_c)( M^{-1}\rd M )^c_b -( M^{-1}\rd M )^a_c (\omega^c{}_b+\Delta^c{}_b)\bigr)h^{eb}\epsilon_{aecd}E^cE^d+\nonumber\\
&&\bigl( \det M \, F f_{ab} E^a E^b + \frac{1}{6} f^{eh} f_{eh} \e_{abcd} E^a E^b E^c E^d \bigr)-\nonumber \\&&
  \frac{2}{3} \bigl( {\nabla} \phi {\phi}^a \e_{abcd} V^b V^c V^d - \frac{1}{8} {\phi}^m {\phi}_m \e_{abcd} V^a V^b V^c V^d \bigr)\Bigr]
 \label{total}\eea
Given the Lagrangian, we can now obtain the equations of motion. The  equations (\ref{gaugefe1},\ref{gaugefe2},\ref{scalarfe},\ref{grav1}) are not modified. Einstein's equations (\ref{grav2}) acquire sources and become the following

\begin{align}
&\partial_c \Delta_a{}^c{}_b - \partial_b \Delta_c{}^c{}_a +
\Delta_c{}^c{}_d \Delta_b{}^d{}_a - \Delta_b{}^c{}_d \Delta_c{}^d{}_a -
F^c{}_{dc} \Delta_b{}^d{}_a - F^d{}_{ab} \Delta_c{}^c{}_d + \nonumber \\
&+ F^c{}_{db} \Delta_c{}^d{}_a + F^d{}_{ac} \Delta_b{}^c{}_d = \frac{1}{2}
F_{xa} h^{ab} F_{yb} - \nabla_x \phi \nabla_y \phi
\end{align}

 \section{Conclusions}

In this paper we have presented a formulation of gravity that allows to use connections with torsion  to describe the gravitational field. It introduces a background, fixed  connection and vielbein. At the global level, there can be an obstruction for the equivalence with the classical formulation, and this allows solutions with topologies that were not possible before.

Implicitly, the formulation has been used in Supergravity compactifications, where the internal manifold has torsion. Torsion has been  called a `geometric flux', and it is a data that together with topological characteristics of the internal manifold determines the theory at low energies. Through T-duality, the torsion or geometric flux in the IIA theory  becomes a standard $3$-form flux \cite{kstt} in IIB. It is known also that T-duality generically changes the topology of the target manifold \cite{aal}, and it is our intention to study   T-duality in the light of this new formulation.

It is not difficult to couple it to fermions and indeed, supersymmetry, as an infinitesimal, local symmetry should be straightforward to implement. We leave it for a future paper, where we will also analyze supergravity compactifications with geometrical fluxes. We want to study if these compactifications are spontaneous or `consistent' from this geometrical point of view, also in the light of previous works \cite{he}.

Finally, and since we have described now a general mechanism, it could be used to find new examples.

\section*{Acknowledgements}
We want to thank very specially to L. Andrianopoli for many important discussions. We also thank to V. S. Varadarajan for his help on the mathematical aspects of the paper.

This work has been supported in part by grants SB2005-0137 of Ministerio de Educaci\'{o}n y Ciencia (Spain), FIS2008-06078-C03-02 of Ministerio de Ciencia e Innovaci\'{o}n (Spain), INFN07-36 and INFN08-13 of Ministerio de Ciencia e Innovaci\'{o}n (Spain) and INFN (Italy) and GVPRE/2008/119 of the Generalitat Valenciana.

\end{document}